\documentclass[preprint,superscriptaddress,article]{revtex4}
\usepackage{amsfonts}
\usepackage{bbm}
\usepackage{mathrsfs}

\usepackage{amsmath,amsthm,amssymb}
\usepackage{graphicx}
\usepackage{dcolumn}
\usepackage{amsmath}
\usepackage{longtable}

\begin{document}

\title{Entropic force and its fluctuation in Euclidian Quantum Gravity}
\author{Yue Zhao }\email{zhaoyue@physics.rutgers.edu}
\affiliation{{\normalsize NHETC and Department of Physics and Astronomy,}\\
{\normalsize Rutgers University, Piscataway, NJ 08854--8019, USA}\\}
\date{\today}

\begin{abstract}
In this paper, we study the idea about gravity as entropic force
proposed by Verlinde.  By interpreting Euclidean gravity in the
language of thermodynamic quantities on holographic screen, we find
the gravitational force can be calculated from the change of entropy
on the screen.  We show normal gravity calculation can be
reinterpreted in the language of thermodynamic variables.  We also
study the fluctuation of the force and find the fluctuation acting
on the point-like particle can never be larger than the expectation
value of the force.  For a black hole in AdS space, by gauge/gravity
duality, the fluctuation may be interpreted as arising from thermal
fluctuation in the boundary description. And for a black hole in
flat space, the ratio between fluctuation and force goes to a
constant $\frac{T}{m}$ at infinity.

\end{abstract}

\maketitle

% Put preprint number in top-right.
\def\pplogo{\vbox{\kern-\headheight\kern -29pt
\halign{##&##\hfil\cr&{\ppnumber}\cr\rule{0pt}{2.5ex}&\ppdate\cr}}}
\makeatletter
\def\ps@firstpage{\ps@empty \def\@oddhead{\hss\pplogo}%
  \let\@evenhead\@oddhead % in case an article starts on a left-hand page
}

% The only change in \maketitle is \thispagestyle{firstpage} instead of \thispagestyle{plain}
\def\maketitle{\par
 \begingroup
 \def\thefootnote{\fnsymbol{footnote}}
 \def\@makefnmark{\hbox{$^{\@thefnmark}$\hss}}
 \if@twocolumn
 \twocolumn[\@maketitle]
 \else \newpage
 \global\@topnum\z@ \@maketitle \fi\thispagestyle{firstpage}\@thanks
 \endgroup
 \setcounter{footnote}{0}
 \let\maketitle\relax
 \let\@maketitle\relax
 \gdef\@thanks{}\gdef\@author{}\gdef\@title{}\let\thanks\relax}
\makeatother

\def\eq#1{Eq. (\ref{eq:#1})}
\def\theequation{\thesection.\arabic{equation}}

%%%%%%%%%%%%%%%%%%%%%%%%%%%%%%%%%%%%%%%%%%%%%%%%%%%%%%%%%%%%%%%%%%%%%%%%%%%%%%%%%%%%%%%%%%%%%%%%%%%%%%%%%%%%%%%%%%%%%%%%%%%%%%%%%%%%%%%%%%%%%%%%%%%%

%\setcounter{page}0
%\def\ppnumber{\vbox{\baselineskip14pt
%\hbox{\footnotesize{RUNHETC-2010-06}} }}
%\def\ppdate{}\date{}
%\author{YZ\\
%[7mm]
%{\normalsize NHETC and Department of Physics and Astronomy,}\\
%{\normalsize Rutgers University, Piscataway, NJ 08854--8019, USA}\\
%}

%\title{\bf Entropic force and its fluctuation from gauge/gravity duality\vskip 0.4cm} \maketitle

\section{Introduction}
The thermodynamics of black hole has been studied for several
decades since the discover of Hawking radiation
\cite{Hawking:1974ur}.  It reveals a deep connection between the
structure and dynamics of space-time and laws of
thermodynamics\cite{summary}.  And more recently, the work by
Jacobson shows an explicit derivation from laws of thermodynamics to
Einstein equation\cite{Jacobson:1995ab}.  The attempts to explain
gravity as an emergent phenomena is based on the holographic
principle \cite{'t Hooft,Susskind} . And AdS/CFT correspondence
provides strong support and explicit examples on how thermodynamics
of space-time can be related to thermodynamics of the dual system
living on holographic
screen\cite{Maldacena:1997re,GKP,Witten,Susskind:1998dq}.

Recently, Verlinde \cite{Verlinde:2010hp} proposed a conjecture that
the origin of gravity can be interpreted as entropy changing on the
holographic screen, which can be explicitly expressed as
\begin{equation}\label{eqn:I1}
F\Delta x=T \Delta S
\end{equation}

There are several assumptions required to realize his idea. Firstly,
one imposes how entropy on elements of holographic screen changes
when one moves the particle $\Delta x$
\begin{equation}\label{eqn:I2}
\Delta S=2\pi k_B \frac{m c}{\hbar}\Delta x
\end{equation}
where the distance between particle and screen element should be
smaller than the Compton wavelength of the particle.  And this will
imply the Unruh-like relation between temperature and acceleration
\begin{equation}\label{eqn:I3}
k_B T=\frac{1}{2\pi}\frac{\hbar a}{c}
\end{equation}
Then to relate energy and temperature of the system, one imposes the
equipartition of energy, see also \cite{Padmanabhan:2009vy}
\begin{equation}\label{eqn:I4}
E=\frac{1}{2}N k_B T
\end{equation}
where $N$ is the number of degree of freedom on the holographic
screen as
\begin{equation}\label{eqn:I5}
N=\frac{A c^3}{G\hbar}
\end{equation}
With identifying the total energy $E$ of the system living on
holographic screen with the bulk energy inside the screen, one can
derive the Newton gravitational force and Einstein equation.

There has been many papers following this
conjecture\cite{Gao:2010fw,Zhang:2010hi,Lee:2010bg,Myung:2010jv,KowalskiGlikman:2010ms,
Liu:2010na,Cai:2010sz,Caravelli:2010be,Tian:2010uy,Myung:2010rz,Vancea:2010vf,Konoplya:2010ak},
and some other papers apply this idea into
cosmology\cite{Cai:2010hk,Li:2010cj,Wei:2010ww,Ling:2010zc},
especially, \cite{Wang:2010jm} studies possible fluctuation seeds
induced by thermal fluctuation on holographic screen during
inflation.

\eqref{eqn:I1} is the equation showing the most important point of
the whole idea, one interprets the origin of gravitational force as
the change of entropy on the holographic screen.  However, to derive
the correct expression for gravitational force or equation, one
needs to impose several other conditions. How far can those imposed
conditions go? In what circumstance do those conditions fail to
work?  Is the interpretation of gravity always correct or just a
coincidence in some particular cases?

In section 2, we apply some basic concepts and relations obtained
from AdS/CFT to study the entropic interpretation of gravity,
leaving those potentially unsafe conditions along, one finds that
only by imposing \eqref{eqn:I1}, one can obtain the correct
expression of gravitational force for generic static spherically
symmetric metric background.  The derivation in this section is an
analog to the calculation in polymer molecule system, which is used
to motive entropic force in Verlinde's paper.

In section 3, we consider our method to derive entropic force in
more details, and we find it has a good interpretation from
gravitational side alone, without correspondence between gravity and
field theory.  The point of this section is following: if one
interprets the derivation from gravity side along, the whole
calculation is just generalized Euler-Lagrange equation. However, if
one gives thermodynamic interpretation ( as quantities on
holographic screen ) to each step during derivation as we did in
section 2, one will automatically draw the conclusion that gravity
as entropic force. Together the analysis in section 2, it provides a
solid base for entropic force interpretation.  Moreover, since our
derivation can be done in a quite generic metric background, it
could be a clue to find more explicit connections and understanding
on thermodynamics of two systems.

In section 4, since the force can be interpreted as a quantity in
thermodynamics, we go further step to study its thermal fluctuation.
We find the thermal fluctuation $\overline{F^2-\overline{F}^2}$ is
always positive. And for a point-like particle, the fluctuation is
never larger than the real force.  The metrics of black hole in
asymptotic flat and AdS space are taken as examples in this section.

\section{Entropic Force in gauge/gravity duality}
One of the main ideas in Verlinde's paper is to interpret the origin
of gravity as the change of the entropy in the dual theory on
holographic plane.  Gauge/gravity correspondence provides us a
natural and very-well understood place to test this idea.  The
change of entropy in field theory on the holographic screen of the
bulk space should give the correct expression for gravitational
force experienced by particles in bulk space.

In gauge/gravity duality, one has the identification between
Euclidian action of the field theory and the Euclidian action from
gravity. And this will be our starting point to see how change of
entropy causes gravity.

Firstly, we consider the simplest case, one point-like particle
moving in a fixed geometric background with black hole.  And we need
to assume that the gravity set-up has a dual theory on field theory
side. The temperature of black hole in gravity side should be
identified as the temperature of the field theory.

Suppose there is an external force acting on the point particle
holding it fixed.  From the boundary point of view, the black hole
plays the role of heat bath, the point particle should be
interpreted as a perturbation away from equilibrium in the bath. If
there is no external force acting on the particle, the particle will
fall into black hole, which is analog to the process that the
perturbation is erased and gets equilibrium with heat bath.  When
there is force holding particle fixed outside horizon, this
corresponds to an effective force keeping perturbation away from
thermal equilibrium with the bath.

Now let us applying the similar analysis used in section 2 in
Verlinde's paper \cite{Verlinde:2010hp} for polymer molecule system.
We choose to use micro-canonical ensemble.  Thus, the entropy of the
whole system can be written as $S(E+F_{ext} r_0,r_0)$, where $E$ is
the total energy of black hole and particle (heat bath and molecule
chain). Since in micro-canonical ensemble one has

\begin{equation}\label{eqn:R1}
\frac{d}{d r_0}S(E+F_{ext}r_0,r_0)=0
\end{equation}

which implies

\begin{equation}\label{eqn:R2}
\frac{F_{ext}}{T}-\frac{\partial S}{\partial r_0}=0
\end{equation}

The next thing to do is to find the expression for entropy from our
gravitational set-up.  To achieve that, one recalls that in
gauge/gravity duality, the Euclidean action calculated from gravity
is identified as entropy of the dual field theory if one treats the
system as micro-canonical ensemble.  That provides a solid way to
calculate entropy in \eqref{eqn:R1}.  The Euclidean action of our
gravitational set-up can be written as
\begin{equation}\label{eqn:R3}
I=I_{Grav}+I{part}+I{int}
\end{equation}
where $I_{Grav}$ is the Euclidean Einstein-Hilbert action of
background metric.  $I_{part}$ is the contribution from the static
particle outside black hole. And $I_{int}$ is from the interaction
which keeps particle fixed.  One can easily see that, in
micro-canonical ensemble, $I_{int}$ will contribute the first term
in \eqref{eqn:R2}, and $(I_{Grav} + I_{part})$ contribute to the
second term.  So to calculate the expression for external force, one
needs to do derivative on $(I_{BG}+I_{part})$ respect to $r_0$.

The action of a point particle in general geometric background is
proportional to the proper mass times the integral of its proper
time
\begin{equation}\label{eqn:P2}
I_{part}={i}m\int d\lambda
\end{equation}
We want to consider the particle fixed at some point.  So one can
write the coordinates of the particle as
$X^\mu=\{t,r_0,\theta_0,\phi_0\}$.

Since we need the Euclidian action, so we take $\lambda\rightarrow
{i}\tau$.  And one has to give the correct range where Euclidian
proper time $\tau$ runs.

To find the range of $\tau$, one firstly recalls how temperature
relates to the metric.  Suppose one has a static spherically
symmetric metric
\begin{equation}\label{eqn:P3}
ds^2=-g_{tt}(r)dt^2+g_{rr}(r)dr^2+...
\end{equation}

Here $g_{tt}$ and $g_{rr}$ are only functions of $r$, and $...$ just
the normal angular parts of the metric.  After taking $t\rightarrow
{i}{t}$, one finds ${t}$ has to have period as
$$\beta=\frac{1}{T}=\frac{4\pi}{\sqrt{g_{tt}'(r)(\frac{1}{g_{rr}(r)})'|_{r=r_h}}}$$
where we set $k_B$ to 1, $T$ is identified as temperature of the
system, and $r_h$ is the position of horizon.

This is the period of coordinate time $t$, and it differs with
particle's proper time $\tau$ by a factor of $\sqrt{g_{tt}}$.  Thus
one has
\begin{equation}\label{eqn:P4}
I_{part}=-m\frac{\sqrt{g_{tt}}}{T}
\end{equation}

Here we neglect the back reaction from the particle to background
metric.  This is corresponding to treat the heat bath, where the
perturbation sits, infinitely large, and the small variation of the
perturbation is infinitesimal respect to whole system.  Then
$I_{Grav}$ actually has no dependence on the position of particle.
Now one can study how entropy changes respect to $r_0$
\begin{equation}\label{eqn:P5}
\Delta S=\frac{\partial I_{part}}{\partial r_0}\Delta
r_0=-\frac{m}{2}\frac{\partial_{r_0} g_{tt}}{\sqrt{g_{tt}}T}\Delta
r_{0}
\end{equation}

Now applying Verlinde's conjecture, taking the origin of
gravitational force as the change of entropy in dual theory, one
has, in a covariant form,
\begin{equation}\label{eqn:P6}
F_a= T \nabla_a S
\end{equation}

Before going into calculation, several points needs to be clear. The
definitions of work and force need to be clarified.  There are two
sets of coordinates describing the system, so one needs to choose
the coordinates very carefully.

To motivate our choices of coordinates, let us briefly review the
definition of work and force in special relativity.  In special
relativity, one also has two sets of coordinates, the proper
$(\tau,\vec{l}  )$ and observer $(t,\vec{x})$ coordinates.  One
defines work as
\begin{equation}\label{eqn:P7}
W=\int_{x_0}^{x_1}\vec{F}\cdot d\vec{x}
\end{equation}
and force is defined as
\begin{equation}\label{eqn:P8}
F^i=\frac{d p^i}{d t}=\sqrt{g_{tt}}\frac{d}{d\tau}(m_0\frac{d
x^i}{d\tau})
\end{equation}
Notice according to definition, force showed up here is $not$ a
4-vector, and the upper subscript $i$ on $F$ is coming from $x^i$.
Now one can write the generalized definition for work in a general
background metric.

With background metric, one also has two sets of coordinates, one
set is used to describe the field theory side $(\tau,\vec{l})$ which
should be interpreted as proper coordinates, another set is used in
gravity theory side $(t,\vec{x})$, which is analog to observer
coordinates in special relativity.  Thus one can write down the
definition of work in general background metric as
\begin{equation}\label{eqn:P9}
W=\int_{x_0}^{x_1}F_i dx^i=\int_{x_0}^{x_1}g_{ij}F^i dx^j
\end{equation}
And in our case, we only care the radial component, thus we have
\begin{equation}\label{eqn:P10}
\Delta W=F_r \Delta r=g_{rr}F^r \Delta
r=g_{rr}\sqrt{g_{tt}}\frac{d}{d\tau}(m_0\frac{d r}{d\tau}) \Delta r
\end{equation}
Now we can plug \eqref{eqn:P5} and \eqref{eqn:P10} into
\eqref{eqn:P6} and find the expression of the force as
\begin{equation}\label{eqn:P11}
m_0\frac{d^2 r}{d\tau^2}=-\frac{m_0}{2}
\frac{1}{g_{tt}g_{rr}}\partial_{r_0}g_{tt}
\end{equation}

Since the particle is static, the external force should be balanced
by gravitational force.  And from \eqref{eqn:R2}, this force emerges
from the changing of entropy of the dual system on holographic
screen.  If the entropic force interpretation is correct, one should
get the same answer as one gets from geodesic equation. From the $r$
component of geodesic equations, one has
\begin{equation}\label{eqn:P11}
\frac{d^2 r}{d \tau^2}+\frac{1}{2}\Gamma^{r}_{\mu\nu}\frac{d
X^{\mu}}{d\tau}\frac{d X^{\nu}}{d\tau}=F_{ext}
\end{equation}
$F_{ext}$, again, is the external force keeping the particle fixed
at $r_0$. Since the particle is not moving on radial direction, and
the external force is balanced by gravitational force, one gets
\begin{equation}\label{eqn:P12}
F_{Grav} =m\frac{d^2 r}{d\tau^2}=-F_{ext}=
-\frac{m}{2}\frac{1}{g_{tt}g_{rr}}\partial_{r_0} g_{tt}
\end{equation}
which is exactly the same expression as the gravitational force
calculated from entropic force interpretation using the
thermodynamic language on holographic screen.

In the derivation, starting from micro-canonical ensemble, $ F_a= T
\nabla_a S$ is the only formula we use, it does not depend on other
assumptions, such as equipartition of energy.  So this is a very
safe check on the idea about the entropic origin of gravity.  And
one can claim that gravity always points to the direction to
increase the system entropy, which leads the phenomena that gravity
is always attractive.

\section{Understanding Entropic Force From Gravity Side Alone}

In previous section we have seen how entropic force interpretation
in dual field theory gives the correct expression for gravity in
bulk. In this section, we will consider our derivation again but
interpret only from gravity side alone, without any knowledge about
dual description on holographic screen.

The way we work in last section is firstly to write down the total
action, then identify it with the entropy in field theory side.
Taking the expression for black hole temperature, and applying
\eqref{eqn:I1}, one can derive the expression for gravity.

Now let us firstly forget about the field theory interpretation, and
only consider the point-like particle in background metric. Instead
of writing the Polyakov point particle action, we write it in
Nambu-Goto form as

\begin{equation}\label{eqn:U1}
I_{part}={i}m\int d\lambda \sqrt{\dot{X_\mu}\dot{X^{\mu}}}
\end{equation}
For a particle fixed at some point, after gauge fixing $\lambda$ as
particle's proper time, the action just reduces to Polyakov point
particle action.

Again, since the particle is in the heat bath by black hole, the
Euclidian time direction, after taking $\lambda\rightarrow {i}\tau$,
should be periodical.

Following the standard procedure in classical mechanics, consider a
system whose action integration range is coordinate dependent
\begin{equation}\label{eqn:U2}
S=\int_0^{f(q)} d\tau (L_{part}[q,\dot{q}]+L_{int}[q])
\end{equation}
here $q$ is the canonical coordinate and $\dot{q}=dq/d\tau$,
$L_{int}$ is the interaction term acting on the particle to keep it
fixed, and it will give the term of $F_{ext}$ in equation of motion.
E.O.M. is given by functional derivative of $q$ and $\dot{q}$. Thus
one gets
\begin{equation}\label{eqn:U3}
\delta S=f(q) \frac{\partial L}{\partial
q}-f(q)\frac{d}{d\tau}\frac{\partial L}{\partial
\dot{q}}+\frac{\partial f(q)}{\partial q}L[q,\dot{q}]=0
\end{equation}
In our case,
\begin{equation}\label{eqn:U4}
L_{part}=m \sqrt{\dot{X_\mu}\dot{X^{\mu}}}
\end{equation}
and $q$ is taken as $r$.  Thus,
\begin{equation}\label{eqn:U5}
\frac{d}{d\tau}\frac{\partial L}{\partial \dot{r}}=g_{rr} \frac{d^2
r}{d\tau^2}
\end{equation}

Taking \eqref{eqn:U5} into \eqref{eqn:U3}, and notice in our case
$f(q)=\frac{\sqrt{g_{tt}}}{T}$, one can again generate
\eqref{eqn:P12}, which is what we expect.

We can see that, the same calculation procedure, from gravity side
point of view without knowledge anything about dual description, is
just solving the generalized Euler-Lagrange equation. The point here
is that the normal gravity calculation can be reinterpreted in the
language of thermodynamic variables on the holographic screen, where
gravity can be treated as change of entropy from dual theory point
of view.  The concept of gravity as entropic force just comes out
naturally if the correspondence between gravity and field theory is
set up and interpreted correctly.

\section{Thermal Fluctuation}
In the previous sections, we checked the entropic interpretation of
gravitational force in the language of thermodynamics variables on
holographic screen. And since we interpret the force as a
thermodynamic quantity, we can push this idea one more step to study
the fluctuation of gravitational force.  In our set-up, $F_{Grav}$
and $r_0$ are analog to pressure and volume in dual field theory
side. And since the particle is held by external force and the
system is in equilrium, one is allowed to use normal thermodynamics
analysis to do calculation.  Since we are interested in the
fluctuation of the force acting on the particle, i.e. the
fluctuation of the gravitational force, we can treat
$(I_{Grav}+I_{part})$ as the partition function of the system.  One
can consistently find that the first derivative of partition
function respect to $r_0$ does give the correct expression for
external force, which is the standard way to calculate pressure in
normal thermodynamic analogy.

Let us first do some analysis on fluctuation in usual statistical
system. Partition function of the system can be written as
\begin{equation}\label{eqn:S1}
Z=\sum_s {e}^{-\beta E_s}
\end{equation}
External force $Y$, by definition, is
\begin{equation}\label{eqn:S2}
Y=\frac{\partial E_s}{\partial y}
\end{equation}
$y$ is the extensive quantity corresponding to $Y$, so
\begin{equation}\label{eqn:S9}
\overline{Y}=\frac{\sum_s \frac{\partial E_s}{\partial
y}{e}^{-\beta(E_s)}}{Z}
\end{equation}
Then the fluctuation of $Y$ can be derived as
\begin{equation}\label{eqn:S10}
\overline{Y^2}-\overline{Y}^2=\frac{1}{\beta^2}\frac{\partial^2}{\partial
y^2} Ln Z +\frac{1}{\beta}\overline{\frac{\partial Y}{\partial y}}
\end{equation}
If the system is adiabatic and in equilibrium, one has
\begin{equation}\label{eqn:S11}
\overline{\frac{\partial^2 E}{\partial
y^2}}=\overline{\frac{\partial Y}{\partial y}}=\frac{\partial
\overline{Y}}{\partial y}
\end{equation}

%Foundations of Statistical Mechanics Volume I: Equilibrium Theory Walter T. Grandy, Jr.

Now, come back to our analysis, we identify $Y$ with $F$ in
\eqref{eqn:P6}, $y$ with $r_0$ and $Ln Z$ with Euclidian action $I$.
Then we have
\begin{equation}\label{eqn:S12}
\overline{F^2}-\overline{F}^2=\frac{1}{\beta^2}\frac{\partial^2}{\partial
r_0^2} I +\frac{1}{\beta}\overline{\frac{\partial F}{\partial r_0}}
\end{equation}
The set-up of our system is a point particle fixed at a point in
some generic metric background, so it is an adiabatic system in
equilibrium.  Thus we can apply \eqref{eqn:S11}.

% S=\frac{m \sqrt{g_{tt}}}{T}+S_0
% 0=dS=\frac{\sqrt{g_{tt}}}{T} dm +\frac{m}{T}\partial_{r_0}\sqrt{g_{tt}}dr_0
% dY= \partial_{r_0}\sqrt{g_{tt}} dm + m \partial^2_{r_0}\sqrt{g_{tt}} dr_0

After some derivations, one gets a simple formula
\begin{equation}\label{eqn:S13}
\overline{F^2}-\overline{F}^2= T m \frac{(\partial_{r_0}
\sqrt{g_{tt}})^2}{\sqrt{g_{tt}}}=\frac{T}{m\sqrt{g_{tt}}}\overline{F}^2
\end{equation}
One can see that \eqref{eqn:S13} is always positive.  We will study
\eqref{eqn:S13} in some more details in following sections.

\subsection{Asymptotic AdS space with black hole}
Since the duality between AdS space with Schwarzschild black hole
and the CFT with finite temperature is one of the most understood
examples in AdS/CFT, we take the explicit metric of that case into
calculation.  And also we take the near horizon limit of the metric.
Thus we have, for example in $AdS_4$
\begin{equation}\label{eqn:S3}
  g_{tt}= (\frac{r^2}{l^2}-\frac{2GM}{r})
\end{equation}
which gives
\begin{equation}\label{eqn:S4}
  \frac{\overline{F^2}-\overline{F}^2}{\overline{F}^2}
  = \frac{T}{m \sqrt{\frac{r^2}{l^2}-\frac{2GM}{r}}}
\end{equation}
Notice that the force we discuss here is not the conventional
gravitational force experienced by particle in AdS space, i.e.
$F_G=m\frac{d^2 r}{d\tau^2}$, they are proportional to each other by
a coefficient as function of background metric at $r_0$.

The particle should be outside of black hole, so $r>\sqrt[3]{2Gl^2
M}$, so ratio is always finite outside the black hole, it will blow
up at black hole horizon and goes to zero when $r$ goes to infinity.
Then one needs to find the point where the ratio becomes one, which
implies the fluctuation is larger than the average value of
gravitational force.

If the point where ratio becomes to one is close to black hole
horizon, then there is nothing to worry about, since the particle
will experience the heat radiation from black hole which naturally
cause large fluctuation.  The only dangerous case is that point is
far away from black hole horizon, which is counter-intuitive and may
potentially contradict with experiments.  Let us talk about this
possibility carefully.

Since we only care about the point where ratio equals to one is far
away from horizon, we can take $r^3\gg 2G l^2 M=r_{BH}^3$, then
\eqref{eqn:S4}  approximately becomes
\begin{equation}\label{eqn:S6}
\frac{\overline{F^2}-\overline{F}^2}{\overline{F}^2}= \frac{T l}{m
r}
\end{equation}
Using the expression for black hole temperature $T\sim\frac{(G
M^2)^{1/3}}{l^{4/3}}$ and the radius of black hole $R_{BH}\sim (G M
l^2)^{1/3}$, one finds
\begin{equation}\label{eqn:S7}
l m_0\sim\frac{l}{\lambda}\sim (\frac{r_{BH}}{r_{\mathcal {O}(1)}})
\end{equation}
where $r_{\mathcal {O}(1)}$ is the radial position where the ratio
becomes order one, and $\lambda$ is the Compton wavelength of the
particle with mass $m_0$.

From \eqref{eqn:S7}, if the equal ratio point is far away from black
hole horizon, which means $r_{\mathcal {O}(1)}\gg r_{BH}$, then it
requires $\lambda \gg l$.  But we know that the particle with
Compton wavelength larger then AdS radius cannot be well
approximately described by a point-like particle.  This is against
our analysis using Polyakov point particle action.  Thus, within a
self-consistent analysis, one finds the fluctuation experienced by a
point like particle can never dominate the real force, and this can
be treated as a consistency check of our analysis.

Also here is another point needs to be clear.  One requires the
Compton wavelength of particle to be smaller than AdS radius to use
point particle as approximation.  This is not inconsistent with the
treatment as neglecting the back-reaction from particle to
background metric.  To get a stable black hole in AdS space, one
needs to take $M_{BH}\gg \frac{1}{l}$.  As long as the mass of
particle is much smaller compared to black hole mass, our
approximation is safe.

\subsection{Asymptotic flat space with black hole}
The result from \eqref{eqn:S4} is very interesting in the case of
flat space with black hole, the ratio between fluctuation and the
value of gravitational force goes to a constant!
\begin{equation}\label{eqn:S14}
  \frac{\overline{F^2}-\overline{F}^2}{\overline{F}^2}
  = \frac{T}{m}
\end{equation}
$m$ again shows up in the dominator.  However, just like the
previous case, the ratio only becomes bigger than one for a particle
whose Compton wavelength larger than $R_{BH}$.  And it is against
our initial set-up.

We recover the units of quantities, we have
\begin{equation}\label{eqn:S15}
  \frac{\overline{F^2}-\overline{F}^2}{\overline{F}^2}
  = \frac{k_B T}{m c^2}
\end{equation}
We see that even when ratio goes to one in \eqref{eqn:S14}, the
fluctuation comparing with the real value of force is still
extremely small.  However this small number sheds light on an
experimental verification on the idea of entropic force.

\section{Discussion}
In this paper, we study Verlinde's conjecture, gravitational force
is induced by the change of entropy of dual field theory on
holographic screen.  Instead of applying other potentially dangerous
assumptions, we take the idea from gauge/gravity duality,
identifying the Euclidian action of gravity and field theory.  We
calculate the action of a point-like particle held fixed in a static
metric background with black hole. Taking black hole's temperature
as the temperature of the dual system, we get the expression of the
force induced by entropy change when we move the particle along
radial direction. The entropic force derived this way agrees with
the gravitational force in such background metric. Thus, this is a
very safe check on the idea about gravity as emergent phenomena.

The key formula \eqref{eqn:I1} is motivated from thermodynamics of
the dual field theory. To give a more clear picture on how
thermodynamics of gravity is related to that of the dual field
theory, we study the gravity interpretation of \eqref{eqn:I1}.  We
did a similar calculation as entropic force in previous section,
however we kept our interpretation in language of gravity side.  We
find $F_a=T \nabla_a S$ is just corresponding to a generalized
Euler-Lagrange equation in gravity side.  This gives an intuitive
answer on why \eqref{eqn:I1} gives the correct formula for
gravitational force, and provide a solid base for entropic force
interpretation.  From this section, if one sticks on the language of
thermodynamics on holographic screen to describe the system, one can
naturally claim gravity is induced by the entropy change on the dual
field theory.

After checking and giving a more explicit explanation on the
entropic force, we take a further step to study the fluctuation of
gravitational force in our formalism.  we find the fluctuation in
that metric is always positive.  We take two widely used metrics, as
examples.  We find that, for a point-like particle, the fluctuation
will never dominate the real value of gravitational force outside
black hole.  And for asymptotic flat metric with black hole, the
ratio between fluctuation and force goes to a constant at infinity.
This phenomena might provide a clue to experimentally test the
concept of entropic force, or even holographic principle.

\section{Acknowledgements}
I would like to thank S. Thomas for very helpful discussions and
suggestion,  and reading the draft. I also want to thank G. Pan for
useful conversations. This research is supported by NHETC in
Department of Physics and Astronomy at Rutgers University, and in
part by DOE grant DE-FG02-96ER40949.


\begin{thebibliography}{99}

%\cite{Verlinde:2010hp}
\bibitem{Verlinde:2010hp}
  E.~P.~Verlinde,
On the Origin of Gravity and the Laws of Newton.
  arXiv:1001.0785 [hep-th].
  %%CITATION = ARXIV:1001.0785;%%



\bibitem{Hawking:1974ur}
   Hawking, S. W. (1974). "Black hole explosions?". Nature 248 (5443):30



\bibitem{summary}
  J.~D.~Bekenstein,
  %``Black holes and entropy,''
  Phys.\ Rev.\  D {\bf 7}, 2333 (1973).
  %%CITATION = PHRVA,D7,2333;%%
  J.~M.~Bardeen, B.~Carter and S.~W.~Hawking,
  %``The Four laws of black hole mechanics,''
  Commun.\ Math.\ Phys.\  {\bf 31}, 161 (1973).
  %%CITATION = CMPHA,31,161;%%
  S.~W.~Hawking,
  %``Particle Creation By Black Holes,''
  Commun.\ Math.\ Phys.\  {\bf 43}, 199 (1975)
  [Erratum-ibid.\  {\bf 46}, 206 (1976)].
  %%CITATION = CMPHA,43,199;%%
  P.~C.~W.~Davies,
  %``Scalar particle production in Schwarzschild and Rindler metrics,''
  J.\ Phys.\ A  {\bf 8}, 609 (1975).
  %%CITATION = JPAGB,A8,609;%%
  W.~G.~Unruh,
  %``Notes on black hole evaporation,''
  Phys.\ Rev.\  D {\bf 14}, 870 (1976).
  %%CITATION = PHRVA,D14,870;%%

%\cite{Jacobson:1995ab}
\bibitem{Jacobson:1995ab}
  T.~Jacobson,
  %``Thermodynamics of space-time: The Einstein equation of state,''
  Phys.\ Rev.\ Lett.\  {\bf 75}, 1260 (1995)
  [arXiv:gr-qc/9504004].
  %%CITATION = PRLTA,75,1260;%%


\bibitem{'t Hooft}
  G.~'t Hooft,
  %``Dimensional reduction in quantum gravity,''
  arXiv:gr-qc/9310026.
  %%CITATION = GR-QC/9310026;%%

\bibitem{Susskind}
  L.~Susskind,
  %``The World As A Hologram,''
  J.\ Math.\ Phys.\  {\bf 36}, 6377 (1995)
  [arXiv:hep-th/9409089].
  %%CITATION = JMAPA,36,6377;%%

%\cite{Maldacena:1997re}
\bibitem{Maldacena:1997re}
  J.~M.~Maldacena,
  %``The large N limit of superconformal field theories and supergravity,''
  Adv.\ Theor.\ Math.\ Phys.\  {\bf 2}, 231 (1998)
  [Int.\ J.\ Theor.\ Phys.\  {\bf 38}, 1113 (1999)]
  [arXiv:hep-th/9711200].
  %%CITATION = IJTPB,38,1113;%%

\bibitem{GKP}
S. S. Gubser, I. R. Klebanov and A. M. Polyakov (1998). "Gauge
theory correlators from non-critical string theory". Physics Letters
B428: 105¨C114


\bibitem{Witten}
Edward Witten (1998). "Anti-de Sitter space and holography".
Advances in Theoretical and Mathematical Physics 2: 253¨C291.
http://arxiv.org/abs/hep-th/9802150.

%\cite{Susskind:1998dq}
\bibitem{Susskind:1998dq}
  L.~Susskind and E.~Witten,
``The holographic bound in anti-de Sitter space''.
  arXiv:hep-th/9805114.
  %%CITATION = HEP-TH/9805114;%%

%\cite{Padmanabhan:2009vy}
\bibitem{Padmanabhan:2009vy}
  T.~Padmanabhan,
``Thermodynamical Aspects of Gravity: New insights''.
  arXiv:0911.5004 [gr-qc].
  %%CITATION = ARXIV:0911.5004;%%




%\cite{Smolin:2010kk}
\bibitem{1Smolin:2010kk}
  L.~Smolin,
``Newtonian gravity in loop quantum gravity''.
  arXiv:1001.3668 [gr-qc].
  %%CITATION = ARXIV:1001.3668;%%

%\cite{Shu:2010nv}
\bibitem{1Shu:2010nv}
  F.~W.~Shu and Y.~Gong,
``Equipartition of energy and the first law of thermodynamics at the
apparent horizon''.
  arXiv:1001.3237 [gr-qc].
  %%CITATION = ARXIV:1001.3237;%%






%\cite{Gao:2010fw}
\bibitem{Gao:2010fw}
  C.~Gao,
``Modified Entropic Force''.
  arXiv:1001.4585 [hep-th].
  %%CITATION = ARXIV:1001.4585;%%

%\cite{Zhang:2010hi}
\bibitem{Zhang:2010hi}
  Y.~Zhang, Y.~g.~Gong and Z.~H.~Zhu,
``Modified gravity emerging from thermodynamics and holographic
principle''.
  arXiv:1001.4677 [hep-th].
  %%CITATION = ARXIV:1001.4677;%%




%\cite{Lee:2010bg}
\bibitem{Lee:2010bg}
  J.~W.~Lee, H.~C.~Kim and J.~Lee,
``Gravity from Quantum Information''.
  arXiv:1001.5445 [hep-th].
  %%CITATION = ARXIV:1001.5445;%%

%\cite{Myung:2010jv}
\bibitem{Myung:2010jv}
  Y.~S.~Myung,
``Entropic force in the presence of black hole''.
  arXiv:1002.0871 [hep-th].
  %%CITATION = ARXIV:1002.0871;%%


%\cite{KowalskiGlikman:2010ms}
\bibitem{KowalskiGlikman:2010ms}
  J.~Kowalski-Glikman,
``A note on gravity, entropy, and BF topological field theory''.
  arXiv:1002.1035 [hep-th].
  %%CITATION = ARXIV:1002.1035;%%

%\cite{Liu:2010na}
\bibitem{Liu:2010na}
  Y.~X.~Liu, Y.~Q.~Wang and S.~W.~Wei,
``Temperature and Energy of 4-dimensional Black Holes from Entropic
Force''.
  arXiv:1002.1062 [hep-th].
  %%CITATION = ARXIV:1002.1062;%%

%\cite{Cai:2010sz}
\bibitem{Cai:2010sz}
  R.~G.~Cai, L.~M.~Cao and N.~Ohta,
``Notes on Entropy Force in General Spherically Symmetric
Spacetimes''.
  arXiv:1002.1136 [hep-th].
  %%CITATION = ARXIV:1002.1136;%%

%\cite{Caravelli:2010be}
\bibitem{Caravelli:2010be}
  F.~Caravelli and L.~Modesto,
  %``Holographic actions from black hole entropy,''
  arXiv:1001.4364 [gr-qc].
  %%CITATION = ARXIV:1001.4364;%%


%\cite{Tian:2010uy}
\bibitem{Tian:2010uy}
  Y.~Tian and X.~Wu,
``Thermodynamics of Black Holes from Equipartition of Energy and
Holography'',
  arXiv:1002.1275 [hep-th].
  %%CITATION = ARXIV:1002.1275;%%

%\cite{Myung:2010rz}
\bibitem{Myung:2010rz}
  Y.~S.~Myung and Y.~W.~Kim,
``Entropic force and entanglement system'',
  arXiv:1002.2292 [hep-th].
  %%CITATION = ARXIV:1002.2292;%%


%\cite{Vancea:2010vf}
\bibitem{Vancea:2010vf}
  I.~V.~Vancea and M.~A.~Santos,
``Entropic Force Law, Emergent Gravity and the Uncertainty
Principle'',
  arXiv:1002.2454 [hep-th].
  %%CITATION = ARXIV:1002.2454;%%

%\cite{Konoplya:2010ak}
\bibitem{Konoplya:2010ak}
  R.~A.~Konoplya,
``Entropic force, holography and thermodynamics for static
space-times'',
  arXiv:1002.2818 [hep-th].
  %%CITATION = ARXIV:1002.2818;%%




















%\cite{Cai:2010hk}
\bibitem{Cai:2010hk}
  R.~G.~Cai, L.~M.~Cao and N.~Ohta,
``Friedmann Equations from Entropic Force''.
  arXiv:1001.3470 [hep-th].
  %%CITATION = ARXIV:1001.3470;%%


%\cite{Li:2010cj}
\bibitem{Li:2010cj}
  M.~Li and Y.~Wang,
``Quantum UV/IR Relations and Holographic Dark Energy from Entropic
Force''.
  arXiv:1001.4466 [hep-th].
  %%CITATION = ARXIV:1001.4466;%%



%\cite{Wei:2010ww}
\bibitem{Wei:2010ww}
  S.~W.~Wei, Y.~X.~Liu and Y.~Q.~Wang,
``Friedmann equation of FRW universe in deformed Horava-Lifshitz
gravity from entropic force''.
  arXiv:1001.5238 [hep-th].
  %%CITATION = ARXIV:1001.5238;%%

%\cite{Ling:2010zc}
\bibitem{Ling:2010zc}
  Y.~Ling and J.~P.~Wu,
``A note on entropic force and brane cosmology''.
  arXiv:1001.5324 [hep-th].
  %%CITATION = ARXIV:1001.5324;%%


%\cite{Wang:2010jm}
\bibitem{Wang:2010jm}
  Y.~Wang,
``Towards a Holographic Description of Inflation and Generation of
Fluctuations from Thermodynamics''.
  arXiv:1001.4786 [hep-th].
  %%CITATION = ARXIV:1001.4786;%%



\end{thebibliography}
\end{document}